\documentclass[aps,twocolumn,letterpaper,showpacs,longbibliography]{revtex4-1}

\usepackage{amsmath}
\usepackage{graphicx}

\newcommand{\cosps}{\cos^2\varphi}
\newcommand{\sinps}{\sin^2\varphi}
\newcommand{\costp}{\cos(2\varphi)}

\newcommand{\cosp}{\cos\varphi}
\newcommand{\sinp}{\sin\varphi}
\newcommand{\ahat}{\hat{a}}
\newcommand{\adaghat}{\hat{a}^{\dagger}}

\def\bra#1{\mathinner{\langle{#1}|}}
\def\ket#1{\mathinner{|{#1}\rangle}}

\begin{document}
% Title Page
\title{Magnetic properties of the $\alpha$-$T_3$ model: magneto-optical conductivity and the Hofstadter butterfly}
\author{E. Illes$^1$}
\email{illese@uoguelph.ca}
\author{E. J. Nicol$^1$}
\affiliation{$^1$Department of Physics, University of Guelph, Guelph, Ontario N1G 2W1, Canada and\\
Guelph-Waterloo Physics Institute, University of Guelph, Guelph, Ontario N1G 2W1, Canada}%\\
\pacs{78.67.Wj, 78.20.Ls, 72.80.Vp, 72.80.Vp}
\date{\today}

\begin{abstract}
The $\alpha$-$T_3$ model extrapolates between the pseudospin $S=1/2$ honeycomb lattice of graphene and the pseudospin $S=1$ dice lattice via parameter $\alpha$.  We present calculations of the magnetic properties of this hybrid pseudospin model, namely the absorptive magneto-optical conductivity and the Hofstadter butterfly spectra.  In the magneto-optics curves, signatures of the hybrid system corollary a doublet structure present in the peaks, resulting from differing Landau level energies in the $K$ and $K^{\prime}$ valleys.  In the Hofstadter spectra, we detail the evolution of the Hofstadter butterfly as it changes its periodicity by a factor of three as we vary between the two limiting cases of the $\alpha$-$T_3$ model.  
\end{abstract}

\maketitle
\section{Introduction}
Graphene, first experimentally isolated in 2004~\cite{novoselov:2004}, is a two-dimensional sheet of carbon atoms arranged on a honeycomb lattice (HCL).  Its low-energy excitations are described by the two-dimensional massless Dirac equation, or the Dirac-Weyl equation with pseudospin $S=1/2$.  In a magnetic field perpendicular to the lattice, the states of graphene condense into Landau levels (LLs) with energies proportional to $\sqrt{B}$ for both electrons and holes~\cite{mclure:1956}.  These LLs include a zero-energy LL with both electron and hole character, resulting in a half-integer anomalous Hall effect~\cite{zheng:2002,gusynin:2005,zhang:2005}.  

Modifying the HCL by coupling one of the two inequivalent sites of the HCL to an additional atom located at the center of each hexagon yields the $T_3$ or dice lattice~\cite{vidal:2001,vidal:1998,dora:2011}.  This lattice could be naturally formed by growing a tri-layer structure of cubic lattices such as SrTiO$_3$/SrIrO$_3$/SrTiO$_3$ in the $(111)$ direction~\cite{wang:2011} or by confining cold atoms to an optical lattice~\cite{bercioux:2009}.  The low-energy behaviour of the dice lattice is described by the same Dirac-Weyl Hamiltonian as graphene, but with pseudospin $S=1$.  

Allowing a parameter $\alpha$ to describe the strength of the coupling between the HCL and the atom at the center of each hexagon results in the $\alpha$-$T_3$ lattice~\cite{raoux:2014}.  In the limit of $\alpha$ approaching $0$ and $1$, we obtain the HCL (with an inert central atom) and the dice lattice, respectively.  The $\alpha$-$T_3$ model was initially proposed for cold atoms confined to an optical lattice, and more recently, Hg$_{1-x}$Cd$_x$Te in the 2D limit at critical doping has been shown to map onto the $\alpha$-$T_3$ model, with an intermediate value of the coupling parameter $\alpha=1/\sqrt{3}$~\cite{malcolm:2015}.  The $\alpha$-$T_3$ model is characterized by a non-topological Berry phase that varies with the parameter $\alpha$~\cite{louvet:2015}.  In contrast to graphene and the dice lattice, the $\alpha$-$T_3$ model has LLs that form at different energies in the inequivalent $K$ and $K^{\prime}$ valleys~\cite{raoux:2014,illes:2015} for $0<\alpha<1$.

Magneto-optical spectroscopy~\cite{orlita;2010} can be used to probe the underlying electronic structure and excitation spectra by measuring transitions between LLs.  In graphene, it has been used to measure the energy spacing between its unusual LL structure for single~\cite{jiang:2007,deacon:2007} and multi-layer graphene~\cite{shadowski:2006,plochocka:2008}, and to measure its electron and hole velocities.  Magneto-optical conductivity has been calculated for single~\cite{ando:1998} and multi-layer~\cite{koshino:2008} graphene as well as the dice lattice, and general pseudospin systems~\cite{malcolm:2014}.  Here, we calculate the magneto-optical conductivity for the hybrid pseudospin system that can be described as a mixture of pseudospin $S=1/2$ and $S=1$.  We discuss the LL structure of the $\alpha$-$T_3$ lattice such as the different LL energies in the $K$ and $K^{\prime}$ valleys, and examine its effects on magneto-optics curves as a function of the parameter $\alpha$, the magnetic field strength, and with changing chemical potential.

%The interplay of two quantizing fields experienced by charged particles moving through a periodic lattice that is subjected to a perpendicular magnetic field results in the Hofstadter butterfly~\cite{hofstadter:1976}.  

Charged particles moving through a periodic lattice that is subjected to a perpendicular magnetic field experience an interplay of two quantizing fields, resulting in the Hofstadter butterfly~\cite{hofstadter:1976}.  In particular, the periodicity of the lattice creates an electrostatic field that quantizes the motion of the charged particles into Bloch bands.  Similarly, a magnetic field, applied perpendicular to the lattice, quantizes the energy of  the electrons into highly degenerate LLs.  When the length scale of these two quantizing fields is on the same order, the Bloch bands and the LLs compete to split the energy spectrum, resulting in a self similar energy spectrum, called the Hofstadter butterfly.   

Experimental observation of Hofstadter butterfly spectra requires finding a system in which the quantizing fields are able to compete on similar length scales using experimentally achievable fields.  Recently, Moire superlattices~\cite{gumbs:2014}, which can be made from twisted graphene~\cite{moon:2012,bistritzer:2011,wang:2012} or by placing graphene on a hexagonal boron nitride substrate~\cite{dean:2013,yu:2014,ponomarenko:2013,hunt:2013}, have offered this possibility in laboratory achievable fields.  Cold atoms in an optical lattice have also been explored for this purpose~\cite{aidelsburger:2013,miyake:2013}.  

Hofstadter butterfly spectra have been calculated for the HCL~\cite{rammal:1985,claro:1979,gumbs:1997,kohmoto:2006} and the dice lattice~\cite{vidal:1998}, which are the two limiting cases of the $\alpha$-$T_3$ model.  Here, we detail the continous evolution of the Hofstadter butterfly spectrum between these two limiting cases, and provide the difference equation required for calculating Hofstadter butterfly spectra for this intermediate regime.    

The remainder of this paper is laid out as follows.  In section~\ref{sec:model} we describe the $\alpha$-$T_3$ model, including the Hamiltonian and wave functions of the model under a perpendicular magnetic field.  In section~\ref{sec:magneto-optics} we present magneto-optical conductivity curves for the $\alpha$-$T_3$ model and highlight signatures of the hybrid pseudospin system.  Section~\ref{sec:butterfly} contains the difference equation required for calculating Hofstadter butterfly spectra for intermediate values of $\alpha$ and some representative spectra for this regime.  Finally, our conclusions can be found in section~\ref{sec:conclusion}.

\section{The $\alpha$-$T_3$ Model}\label{sec:model}
\begin{figure}[htb]
\centering
%\vspace{0.15cm}
\includegraphics[width=\columnwidth]{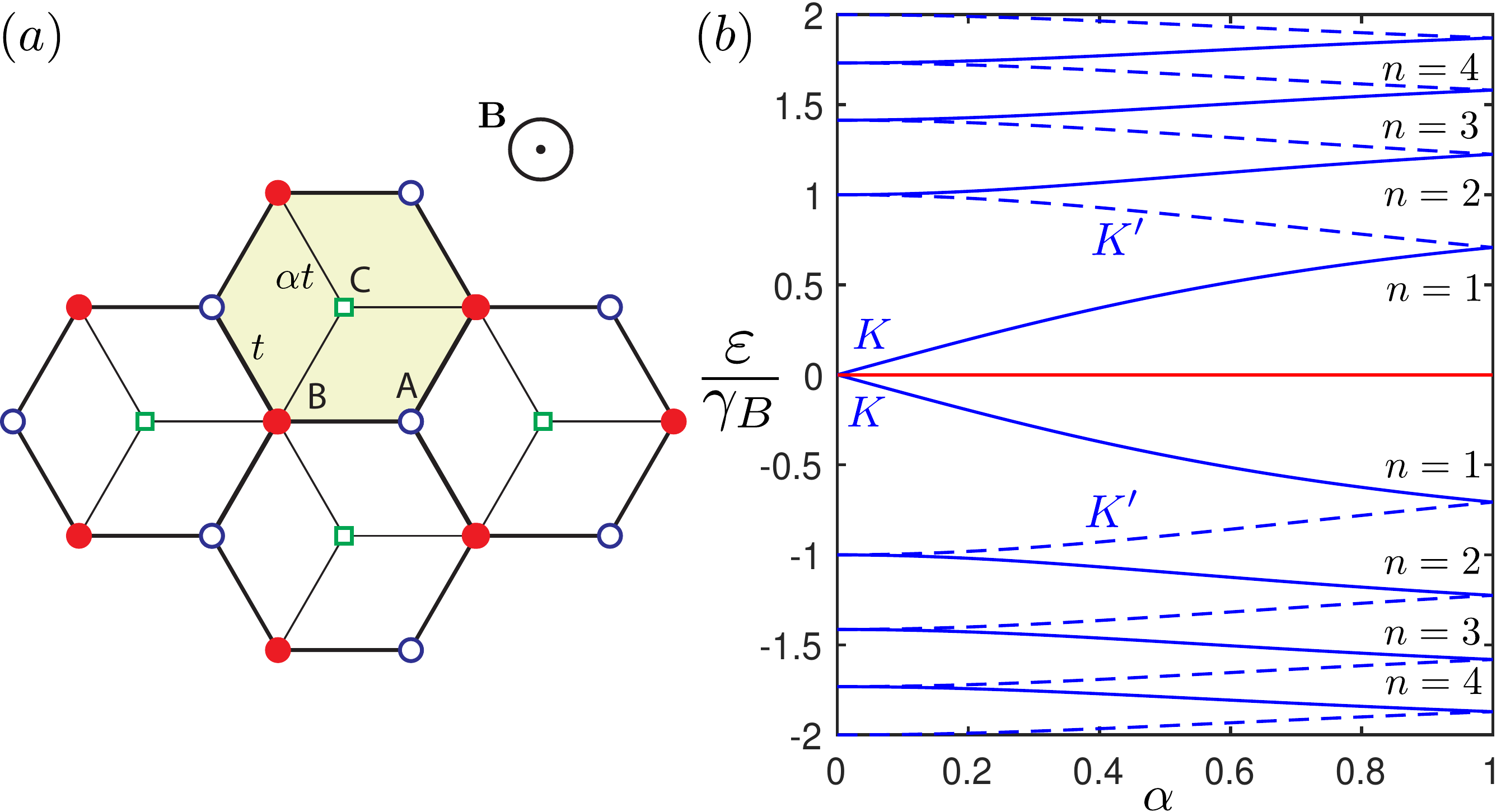}
\caption{(Color online) (a) The $\alpha$-$T_3$ lattice, in a perpendicular magnetic field $B$.  Hopping between sites $A$ and $B$ (which form a HCL) takes place with strength $t$.  Sites labeled $C$, located at the centers of the hexagons, are coupled only to $B$ sites with variable hopping amplitude $\alpha t$. (b) Landau level energies in units of $\gamma_B$ as a function of the parameter $\alpha$ for the first four values of $n$.  The $K$ and $K^{\prime}$ valleys are shown in solid and dashed blue, respectively.  The Landau levels of the flat-band are plotted in red.  }
\label{fig:landau}
\end{figure}

The $\alpha$-$T_3$ model~\cite{raoux:2014} interpolates between the pseudospin $S=1/2$ HCL of graphene, and the pseudospin $S=1$ dice (or $T_3$) lattice via parameter $\alpha$.  Figure~\ref{fig:landau} (a) depicts the $\alpha$-$T_3$ lattice in which sites $A$ and $B$ form a hexagonal lattice, and site $C$ sits at the center of the hexagons.  Hopping takes place between atoms at sites $A$ and $B$ with strength $t$, and a variable hopping of $\alpha t$ connects the $B$ and $C$ sites.  Hopping between sites $A$ and $C$ is not permitted for this model.

Throughout this paper, we will refer to the limiting case of $\alpha=1$ as the dice lattice, and $\alpha\rightarrow 0$ as graphene, for convenience, despite some differences between graphene and the latter limit.  These differences arise from the presence of the $C$ sites, which are located at the center of each hexagon even when they are fully decoupled from the HCL (as is the case for $\alpha=0$).  The result is a three atom per unit cell problem with an inert central atom, rather than the usual two atom per unit cell problem of graphene.  The intermediate regime, in which $0<\alpha<1$, describes a hybrid pseudospin $S=1/2$ and pseudospin $S=1$ system. 

The low-energy spectrum for the $\alpha$-$T_3$ lattice consists of the usual linearly dispersing conical bands expected for graphene, with an additional dispersionless flat-band that cuts through the Dirac point.  All of these bands are present and remain unchanged for the full range of $\alpha$.

In this paper, we are interested in the properties of the $\alpha$-$T_3$ model in the presence of a magnetic field $B$ that is applied perpendicular to the plane of the crystal lattice.  For this case, the low-energy Hamiltonian~\cite{raoux:2014} takes the form
\begin{equation}
H_K= -H^*_{K^{\prime}}= \gamma_B \left(\begin{array}{c c c} 0 & \cosp \ahat & 0 \\
\cosp \adaghat & 0 & \sinp \ahat \\ 
0 & \sinp \adaghat & 0 \\
\end{array}\right)
\end{equation}
with $\gamma_B$ a magnetic energy scale given by $\gamma_B=v_F\sqrt{2eB\hbar}$.  Here $\adaghat$ and $\ahat$ are the creation and annihilation operators, respectively, that obey the usual commutation relation $[\ahat,\adaghat]=1$ and act on Fock states such that $\adaghat\ket{n}=\sqrt{n+1}\ket{n+1}$ and $\ahat\ket{n}=\sqrt{n}\ket{n-1}$.  Note that $\alpha$ has been parametrized by $\alpha=\tan\varphi$ and the Hamiltonian has been scaled by $\cos\varphi$ for convenience~\cite{raoux:2014}.

In the presence of the magnetic field, the electronic states of the $\alpha$-$T_3$ model condense into Landau levels (LLs).  The dispersionless flat-band has zero energy LLs with energy $\varepsilon_{n,0}=0$ for $n=0,2,3,...$.  For the conduction and valence band we have
\begin{align}\label{eq:energy1}
\varepsilon_{n,\pm}&= \pm \gamma_B \sqrt{n-\frac{1}{2}-\frac{\xi}{2}\left(\frac{1-\alpha^2}{1+\alpha^2}\right)}
\end{align}
with $n=1,2,3,...$ and $\xi=\pm$ a valley index for the $K$ and $K^{\prime}$ valley, respectively.  Figure~\ref{fig:landau}(b) depicts the LL structure of the model, as a function of the parameter $\alpha$.  Note the notational difference between the indices of the $\alpha$-$T_3$ model and that of graphene.  Here, the indexing begins with $n=1$ for the conduction and valence band, in contrast to graphene, where it typically starts with $n=0$.

The wavefunctions for the conduction and valence bands for the lowest state $(n=1)$ are
\begin{equation}
\ket{\Psi_{\pm,1}^K} =\frac{1}{\sqrt{2}}\left(\begin{array}{r l}0  & \\
 \pm & \ket{0}\\ 
 & \ket{1}\\
\end{array} \right),
%\end{equation}
%\begin{equation}
\ket{\Psi_{\pm,1}^{K^{\prime}}} =\frac{1}{\sqrt{2}}\left(\begin{array}{r l}  & \ket{1}\\
 \pm & \ket{0}\\ 
0 & \\
\end{array} \right)
\end{equation}
and
\begin{equation}
\ket{\Psi_{\pm,n}^K} =\frac{1}{\sqrt{2}}\left(\begin{array}{r l} \sqrt{\frac{(n-1)\cosps}{n-\cosps}} & \ket{n-2}\\
 \pm & \ket{n-1}\\ 
\sqrt{\frac{n\sinps}{n-\cosps}} & \ket{n}\\
\end{array} \right)
\end{equation}
\begin{equation}
\ket{\Psi_{\pm,n}^{K^{\prime}}} =\frac{1}{\sqrt{2}}\left(\begin{array}{r l} -\sqrt{\frac{n\cosps}{n-\sinps}} & \ket{n}\\
 \pm & \ket{n-1}\\ 
-\sqrt{\frac{(n-1)\sinps}{n-\sinps}} & \ket{n-2}\\
\end{array} \right)
\end{equation}
in general with $n=2,3,4,...$ for the $K$ and $K^{\prime}$ valleys, respectively.    For the flat-band, they are 
\begin{equation}
\ket{\Psi_{0,0}^K} =\left(\begin{array}{r l} 0 & \\
 0 & \\ 
\mp & \ket{0}\\
\end{array} \right),
%\end{equation}
%\begin{equation}
\ket{\Psi_{0,0}^{K^{\prime}}} =\left(\begin{array}{r l} \pm & \ket{0}\\
 0 & \\ 
0 & \\
\end{array} \right)
\end{equation}
for $n=0$ and
\begin{equation}
\ket{\Psi_{0,n}^K} =\left(\begin{array}{r l} \pm\sqrt{\frac{n\sinps}{n-\cosps}} & \ket{n-2}\\
 0 & \ket{n-1}\\ 
\mp\sqrt{\frac{(n-1)\cosps}{n-\cosps}} & \ket{n}\\
\end{array} \right)
\end{equation}
\begin{equation}
\ket{\Psi_{0,n}^{K^{\prime}}} =\left(\begin{array}{r l} \pm\sqrt{\frac{(n-1)\sinps}{n-\sinps}} & \ket{n}\\
 0 & \ket{n-1}\\ 
\mp\sqrt{\frac{n\cosps}{n-\sinps}} & \ket{n-2}\\
\end{array} \right)
\end{equation}
for $n\geq2$.
	
\section{Magneto-Optics}\label{sec:magneto-optics}
\begin{figure}[htb]
\centering
%\vspace{0.15cm}
\includegraphics[width=\columnwidth]{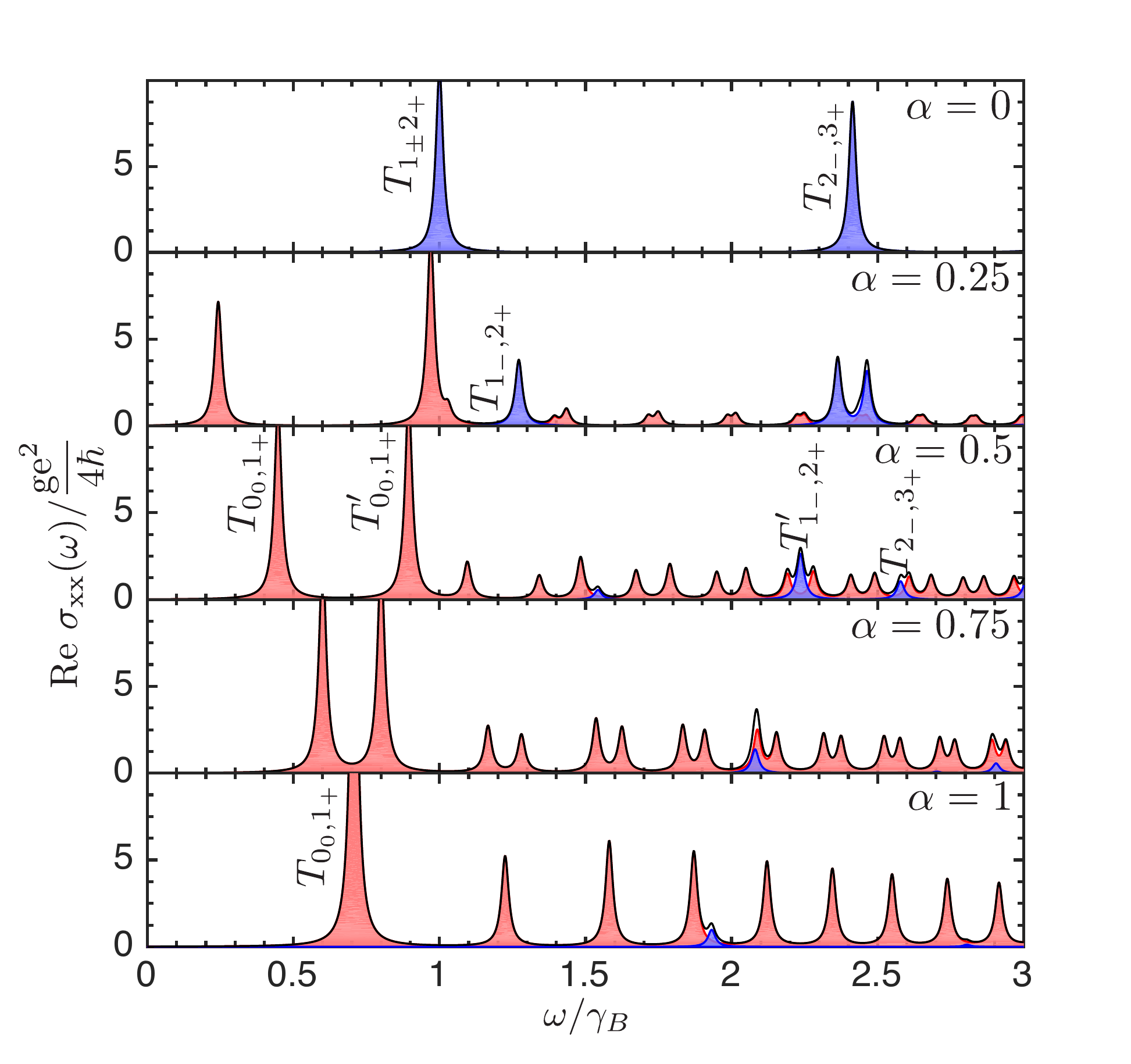}
\caption{(Color online) Absorptive, longitudinal component of the optical conductivity for $\alpha=0,\,0.25,\,0.5,\,0.75,\,1$, from top to bottom, respectively.  The flat-band-to-cone contributions are shaded red, while the cone-to-cone contributions are shaded blue.  Their sum is shown with a thin black curve.  Calculations are done using a scattering rate of $\Gamma=0.025\gamma_B$ and a chemical potential of $\mu=0.1\gamma_B$, which falls below the first positive valued Landau level for all values of $\alpha$ considered.}
\label{fig:sigmaxx_alphas}
\end{figure}

\begin{figure}[htb]
\centering
%\vspace{0.15cm}
 \includegraphics*[width=\columnwidth]{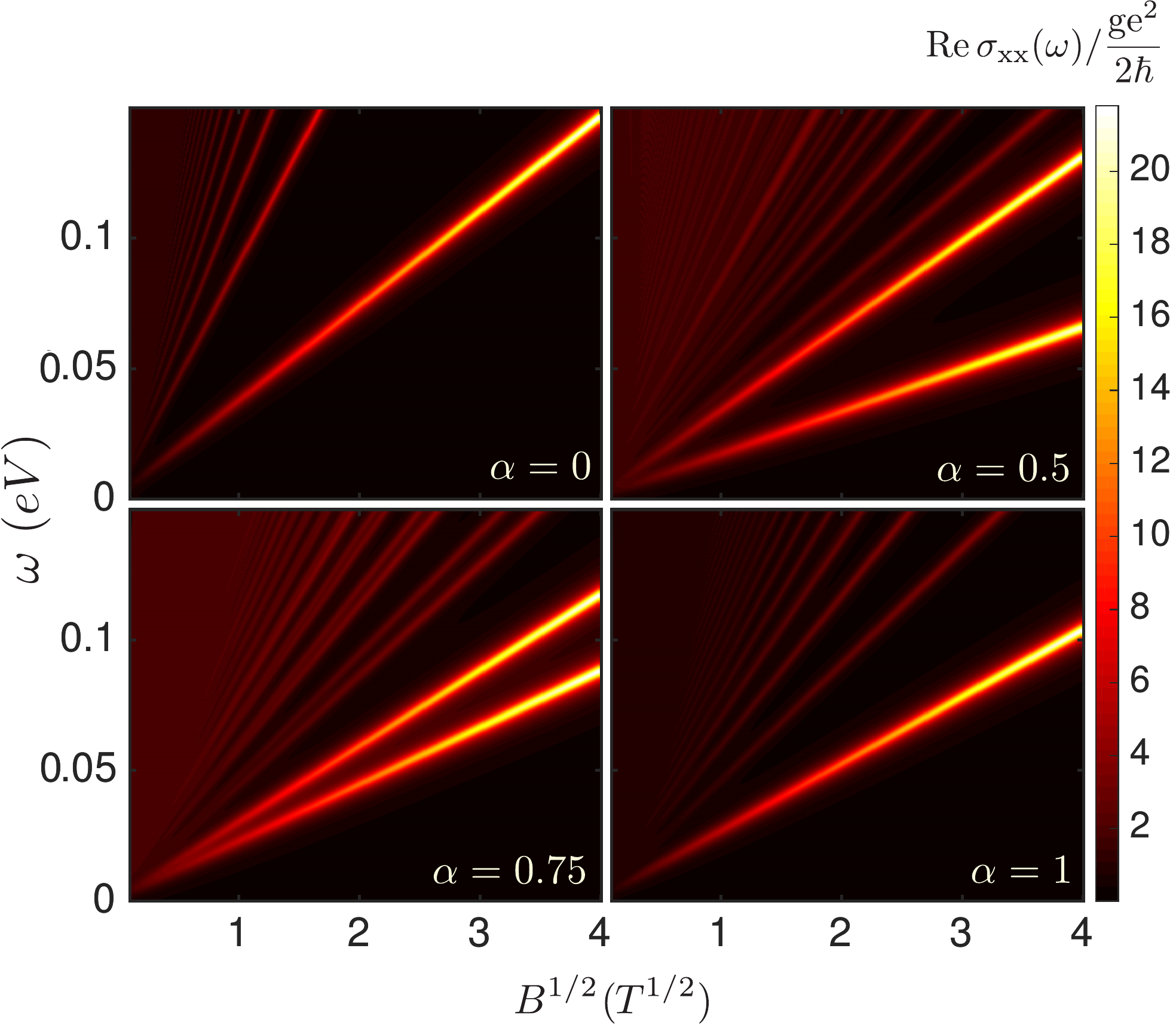}
\caption{(Color online) Absorptive, longitudinal component of the optical conductivity under magnetic fields up to 20T, for $\alpha=0,\,0.5,\,0.75,\,1$.  We used a scattering rate of $\Gamma=2.5$ meV and a chemical potential that falls below the first positive valued Landau level for all values of $\alpha$ considered.}
\label{fig:mag_field_alphas}
\end{figure}
%$\mu=1x10^{-4}eV$ for fig:mag_field_alphas

The absorptive part of the magneto-optical conductivity of a system can be calculated using the Kubo formula.  In the Landau level basis, the Kubo formula can be written
\begin{align}\label{eq:kubo}
	\sigma_{\alpha\beta}(\omega)&=\frac{ig}{2\pi\hbar  l_B^2}\sum_{LLs}\frac{f-f^{\prime}}{\varepsilon^{\prime}-\varepsilon}\frac{\bra{\Psi}{j}_\alpha\ket{\Psi^{\prime}}\bra{\Psi^{\prime}}{j}_\beta\ket{\Psi}}{\omega-(\varepsilon^{\prime}-\varepsilon)+i\Gamma}.
\end{align}	
where $\alpha,\beta=\{x,y\}$ and the summation is over all initial (unprimed) and final (primed) LLs with energy $\varepsilon$ and wavefunction $\ket{\Psi}$.  Here, $f$ is the Fermi factor, $\mu$ is the chemical potential, and $\omega=h\nu$ is the photon energy.  Here, $\Gamma$ can be viewed as the scattering rate of charge carriers, $g$ is the spin degeneracy and $l_B=\sqrt{\frac{hc}{e|B|}}$ is a magnetic length scale.  The current operator is given by ${j}_{\alpha}=-ev_FS_{\alpha}$ with 
\begin{equation}
S_x =\xi\left(\begin{array}{c c c}  0 & \cosp & 0\\
  \cosp & 0 & \sinp \\ 
 0 & \sinp & 0\\
\end{array} \right) 
\end{equation}
\begin{equation}
 S_y =-i\left(\begin{array}{c c c}  0 & \cosp & 0\\
  -\cosp & 0 & \sinp \\ 
 0 & -\sinp & 0\\
\end{array} \right).
\end{equation}	

In the limit of zero temperature and zero scattering rate, the Fermi function $f$ can be written as a Heaviside function $\theta(\mu-\varepsilon)$.  (Additionally, 
$\rm{Im}\frac{1}{\omega-(\varepsilon^{\prime}-\varepsilon)+ i\Gamma}\rightarrow -i\pi\delta(\omega-(\varepsilon^{\prime}-\varepsilon))$ for $\Gamma\rightarrow 0$).
In our magneto-optics calculations these delta functions are broadened by scattering $\Gamma$ as $\delta(x)=\frac{1}{\pi}\frac{\Gamma}{x^2+\Gamma^2}$ where we use $\Gamma$ on the order of $0.025\gamma_B$.	

In order to utilize Eq.~\eqref{eq:kubo} to calculate the magneto-optical response of the system, we require transition matrix elements that describe the probabilities of transitions between LLs.  These can be written  
\begin{flalign}\label{eq:matrix_elements}
		\bra{\Psi_{s,n}^{\xi}}&S_{x}\ket{\Psi_{s^{\prime},n^{\prime}}^{\xi}}\bra{\Psi_{s^{\prime},n^{\prime}}^{\xi}}S_{x}\ket{\Psi_{s,n}^{\xi}}& \nonumber\\
			=&f_1^{\xi,n,n^{\prime},s,s^{\prime}}\delta_{n^{\prime},n+1}+f_2^{\xi,n,n^{\prime},s,s^{\prime}}\delta_{n^{\prime},n-1}\nonumber\\
			\bra{\Psi_{s,n}^{\xi}}&S_{x}\ket{\Psi_{s^{\prime},n^{\prime}}^{\xi}}\bra{\Psi_{s^{\prime},n^{\prime}}^{\xi}}S_{y}\ket{\Psi_{s,n}^{\xi}}& \\
				=&{\xi}f_1^{\xi,n,n^{\prime},s,s^{\prime}}\delta_{n^{\prime},n+1}-{\xi}f_2^{\xi,n,n^{\prime},s,s^{\prime}}\delta_{n^{\prime},n-1}\nonumber		
\end{flalign}	
where $f_1^{\xi,n,n^{\prime},s,s^{\prime}}$ and $f_2^{\xi,n,n^{\prime},s,s^{\prime}}$ are overlap functions between initial (unprimed) and final (primed) states where $\xi=\pm$ for the $K$ and $K^{\prime}$ valleys, respectively; $s=\pm1,0$ for the conduction, valence and flat-band, respectively; and $n$, $n^{\prime}$ is the LL index.  The overlap functions can be written
\begin{align}
	f_1^{\xi,n,n^{\prime},s,s^{\prime}}&=\frac{ n}{4g_1(n)}\left[g_2(n)+2ss^{\prime}C\sqrt{g_1(n)}\right]\nonumber\\
	f_2^{\xi,n,n^{\prime},s,s^{\prime}}&=\frac{ n^{\prime}}{4g_1(n^{\prime})}\left[g_2(n^{\prime})+2ss^{\prime}C\sqrt{g_1(n^{\prime})}\right]\nonumber\\
	f_1^{\xi,n,n^{\prime},s,0}&=\frac{C}{2}\frac{n+1}{n+\frac{1}{2}-\frac{\cos(2\varphi)}{2}}\nonumber\\
	f_2^{\xi,n,n^{\prime},s,0}&=\frac{C}{2}\frac{n-2}{n-\frac{3}{2}-\frac{\cos(2\varphi)}{2}}\nonumber\\
	f_1^{\xi,n,n^{\prime},0,s^{\prime}}&=\frac{C}{2}\frac{n^{\prime}-2}{n^{\prime}-\frac{3}{2}-\frac{\cos(2\varphi)}{2}}\nonumber\\
	f_2^{\xi,n,n^{\prime},0,s^{\prime}}&=\frac{C}{2}\frac{n^{\prime}+1}{n^{\prime}+\frac{1}{2}-\frac{\cos(2\varphi)}{2}}\nonumber\\
	f_{1,2}^{\xi,n,n^{\prime},0,0}&=0
\end{align}	
where $g_1(n)=n^2-\xi n\costp -C$, $g_2(n)=n(1-2C)+\xi\costp(C-1)$, $C=\sinps\cosps$ and $s,s^{\prime}=\pm$, with all cases of $s,s^{\prime}=0$ explicitly shown.  For $\alpha=0,1$ we recover the overlap functions for graphene and the dice model, respectively.

The absorptive diagonal component of the optical conductivity, $\rm{Re}\,\sigma_{xx}(\omega)$ and the absorptive off-diagonal component of the optical conductivity, $\rm{Im}\,\sigma_{xy}(\omega)$ can be calculated from the Kubo formula in Eq.~\eqref{eq:kubo}.  Additionally, for right and left hand polarized light, we can calculate the absorptive optical conductivity as $\rm{Re}\,\sigma_{\pm}(\omega)=\rm{Re}\,\sigma_{xx}(\omega)\mp\rm{Im}\,\sigma_{xy}(\omega)$, respectively.

\begin{figure}[t!]
\centering
%\vspace{0.15cm}
 \includegraphics[width=\columnwidth]{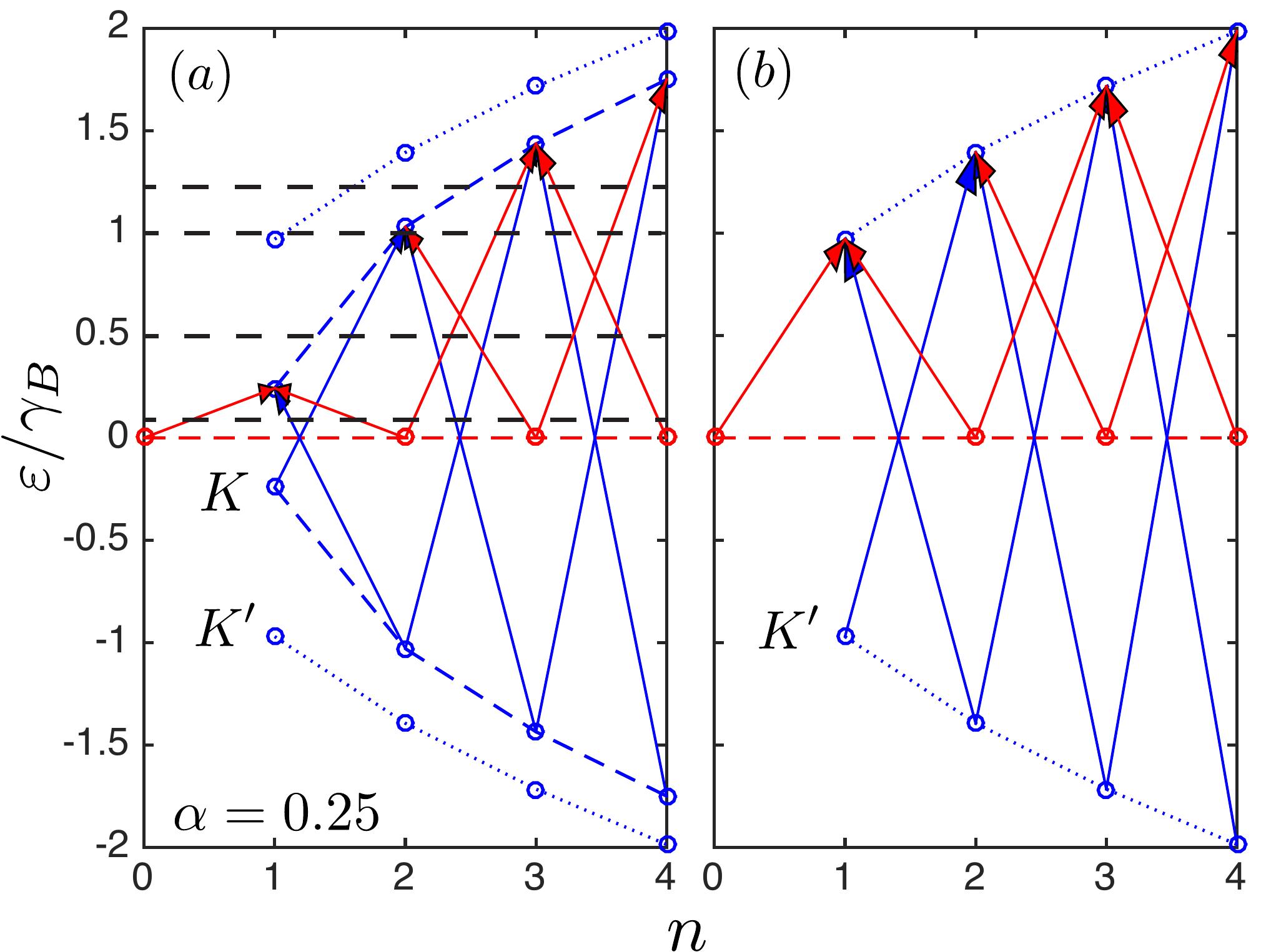}
\caption{(Color online) Snowshoe diagram~\cite{malcolm:2014} for $\alpha=0.25$.  (a) Relative positions of LLs in the $K$ and $K^{\prime}$ valleys are shown with open blue circles connected by dashed and dotted lines, respectively.  Horizontal dashed lines show the four chemical potentials considered in Figs.~\ref{fig:alpha02_sigma_xx} to \ref{fig:alpha02_sigma_minus}.  Arrows represent possible transitions between LLs in the $K$ valley, assuming a chemical potential of $\mu=0.1\gamma_B$.  (b) LL in the $K^{\prime}$ valley with arrows depicting all possible transitions for $\mu=0.1\gamma_B$.}
\label{fig:snowshoe}
\end{figure}

We will use the notation $T_{n_s,n^{\prime}_{s^{\prime}}}$ to denote transitions originating from a LL with index $n_s$ and terminating at a LL with index $n^{\prime}_{s^{\prime}}$.  Primed and unprimed transitions, $T$ and $T^{\prime}$, will denote transitions in the $K$ and $K^{\prime}$ valleys, respectively.  For example, a transition between the first LL of the flat-band ($n=0$) and the first LL in the conduction band ($n=1$) in the $K^{\prime}$ valley would be written $T^{\prime}_{0_0,1_+}$.  Note that for flat-band-to-cone transitions, the energy of the transition $T_{(n+1)_0,n_+}$ is equal to that of $T_{(n-1)_0,n_+}$ and for cone-to-cone transitions, the energy of the transition $T_{(n-1)_-,n_+}$ is equal to that of $T_{n_-,(n-1)_+}$.  For simplicity, we will label peaks resulting from two equal energy transitions using only one of these transitions, unless a distinction needs to be made for our purposes.  Note also that for $\alpha=0,1$, all transitions have the same energy as their primed counterparts (ie, $T_{n,n^{\prime}}=T^{\prime}_{n,n^{\prime}}$).  Additionally, there is no $n=1$ LL for the flat-band, and as a result there are no transitions originating from this LL.

\begin{figure}[t]
\centering
\includegraphics[width=\columnwidth]{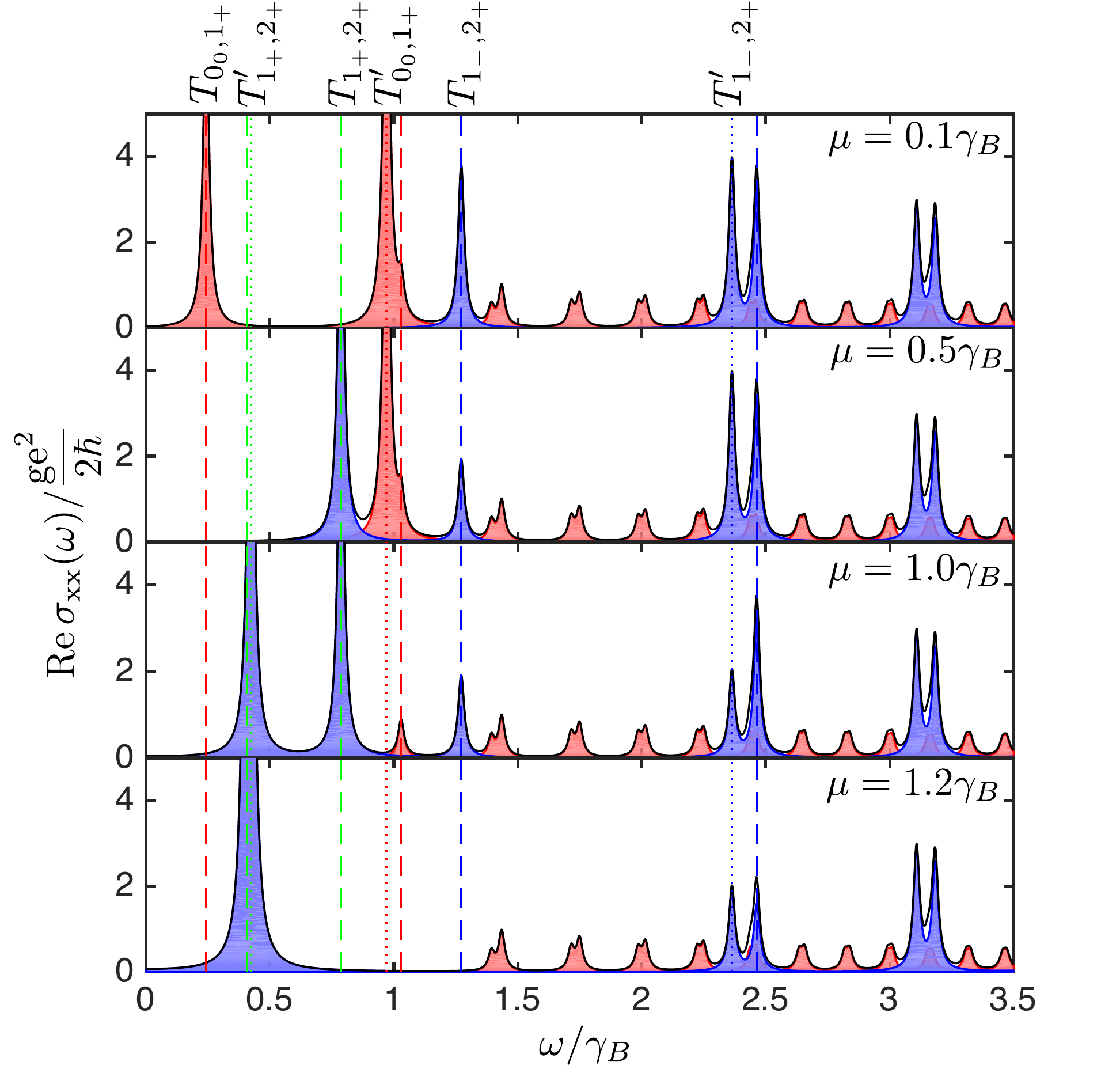}
\caption{(Color online) Absorptive, diagonal component of the optical conductivity for $\alpha=0.25$ showing $\mu/\gamma_B = 0.1$, $0.5$, $1.0$ and $1.2$ from top to bottom, respectively.  The flat-band-to-cone (cone-to-cone) contributions are shaded red (blue), and their sum is represented by a thin black line.  Vertical dashed (dotted) lines show the energies associated with a number of transitions in the $K$ ($K^{\prime}$) valley.  Red, blue and green vertical lines mark the energy of flat-to-cone, cone-to-cone interband, and cone-to-cone intraband transitions, respectively.  In order from left to right, the following transitions are marked: $T_{0_0,1_+}$, $T_{2_+,3_+}$, $T_{1_+,2_+}^{\prime}$, $T_{1_+,2_+}$, $T_{0_0,1_+}^{\prime}$, $T_{1_0,2_+}$, $T_{1_-,2_+}$, $T_{1_-,2_+}^{\prime}$, $T_{2_-,3_+}$.  Only a subset of these are labelled above the plot. }\label{fig:alpha02_sigma_xx}
\end{figure}

In Fig.~\ref{fig:sigmaxx_alphas} we plot the absorptive diagonal component of the optical conductivity for a range of $\alpha$ values.  They are calculated at a chemical potential $\mu=0.1\gamma_B$, such that the smallest positive LL is above the chemical potential, and the zero energy flat-band is below the chemical potential, for all values of $\alpha$ considered.  Cone-to-cone transitions are shaded blue, and flat-band-to-cone transitions are shaded red.  A thin black line shows the total optical response of the system.  Note again that the indexing for the LLs of the conduction and valance band of the $\alpha$-$T_3$ model begin with $n=1$, in contrast to the usual $n=0$ lowest LL of graphene, which results in LL labelling that differs from what is typical for graphene.

For $\alpha=0$ we find only cone-to-cone transitions, as expected for graphene.  For the other limiting case of $\alpha=1$, the flat-band-to-cone transitions dominate, and the cone-to-cone transitions are largely suppressed.  There remains only a comparatively small peak for transition $T_{1_-,2_+}$%/$T_{2_-,1_+}$)
.  This was also noted in Ref~\cite{malcolm:2014} in their magneto-optics calculations for the pseudospin $S=1$ system.    

In the intermediate regime, we observe the coexistence of cone-to-cone, and flat-band-to-cone transitions, and the evolution between the two limiting cases.  This regime is characterized by peaks with anomalous heights or locations that break up the regular pattern of the dominant transition type.  For example, the $T^{\prime}_{1_-,2_+}$ cone-to-cone transition disrupts the dominant pattern of the flat-band-to-cone transitions.  For $\alpha=0.5$ this results in a triplet centered about the $T^{\prime}_{1_-,2_+}$ transition, as its energy corresponds to the center of a flat-band-to-cone doublet.  For $\alpha=0.75$, the presence of the transition lines up with one of the peaks from a flat-band-to-cone doublet, and manifests as an anomalously sized doublet with increased weight on the low energy side.  

\begin{figure}[t]
	\centering
\includegraphics[width=\columnwidth]{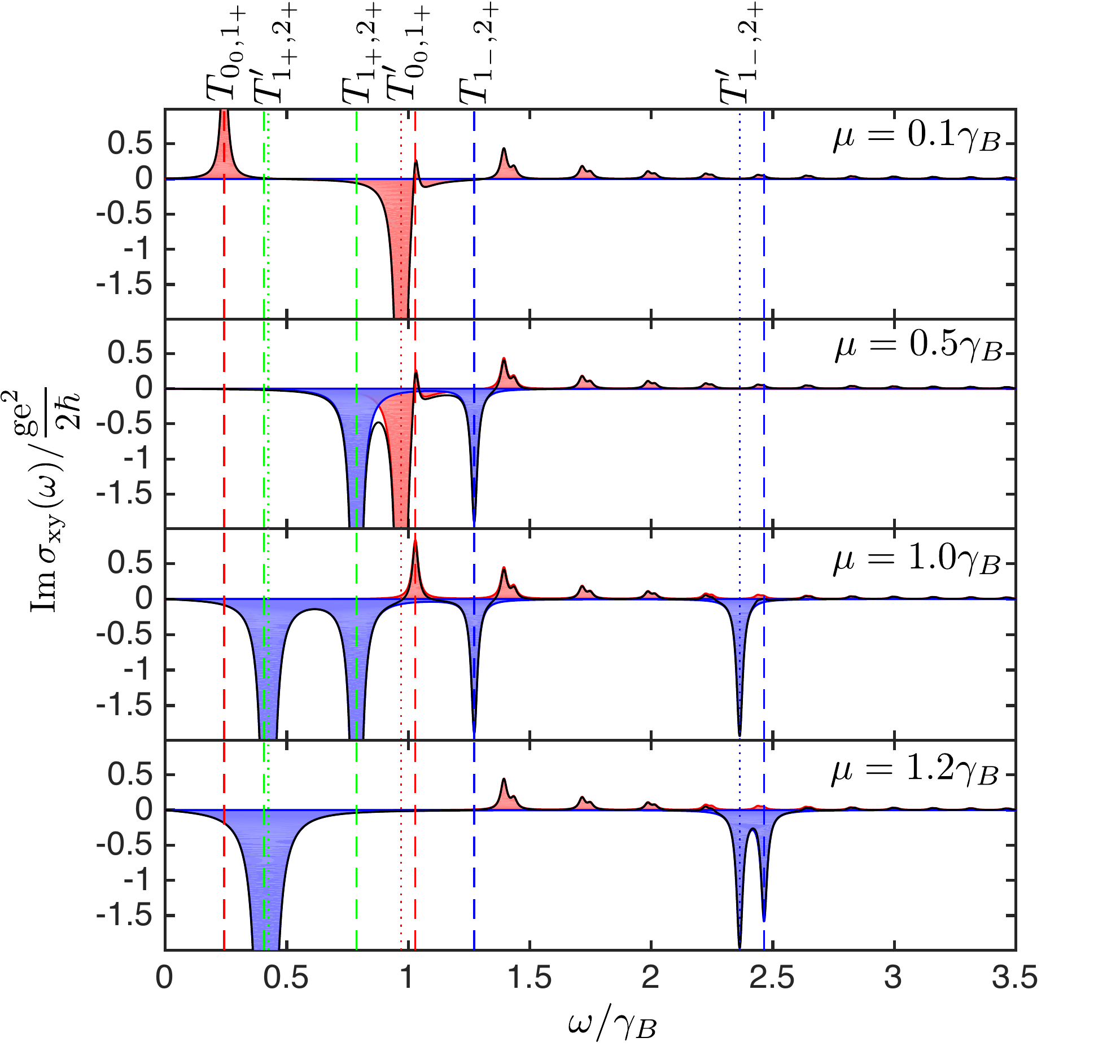}
\caption{(Color online) Same as Fig.~\ref{fig:alpha02_sigma_xx} but for the absorptive, off-diagonal component of the optical conductivity.}\label{fig:alpha02_sigma_xy}
\end{figure}

Additionally, we also note a doubling of the number of peaks in the spectrum in the intermediate regime (both for cone-to-cone and flat-band-to-cone transitions).  This is a consequence of the difference in energies of the LLs in the $K$ and $K^{\prime}$ valleys.  For values of $\alpha$ close to $1$, the doublets observed in the conductivity curves are formed by transitions with identical indices in the $K$ and $K^{\prime}$ valley.  For example $T_{0_0,1_+}$ and $T^{\prime}_{0_0,1_+}$ are flat-band-to-cone doublets present for the full range of $\alpha$ in the intermediate regime, with varying separation between them.  

In the other limit, for $\alpha$ close to $0$, the doublets are formed by transitions with indices that differ by one in the two valleys.  For example, $T^{\prime}_{1_-,2_+}$ forms a doublet with $T_{2_-,3_+}$ in that limit.  This is nicely illustrated in the LL diagram in Fig.~\ref{fig:landau}(b), where we see that for values of $\alpha$ near $1$, the LLs with the same index come together from the two valleys, whereas in the other limit, LLs with indices that differ by one converge ($n+1$ from the $K$ valley meets $n$ from the $K^{\prime}$ valley).

\begin{figure*}
\centering
\begin{minipage}[t]{.45\textwidth} 
	\includegraphics[width=\columnwidth]{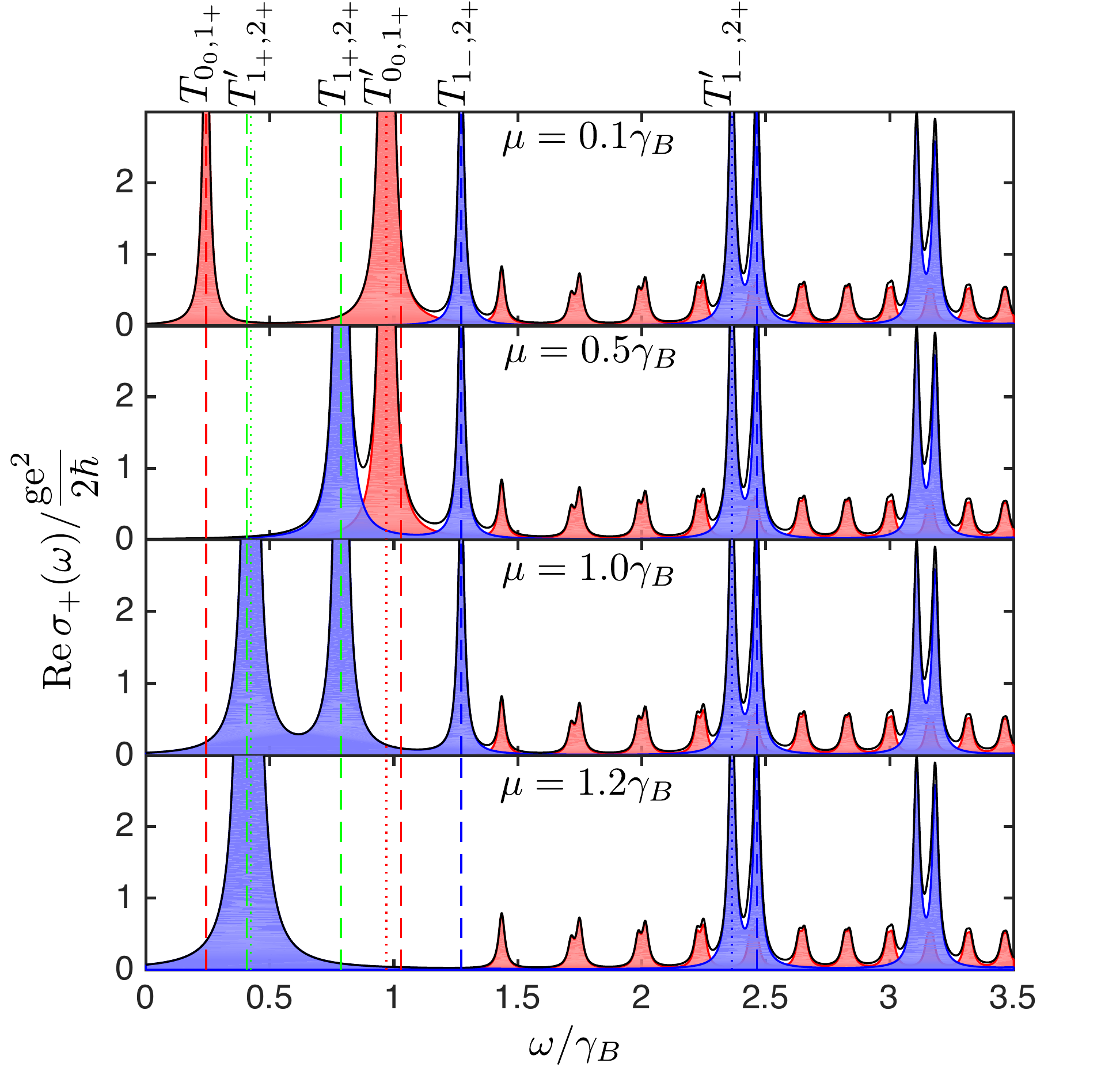}
\caption{(Color online) Same as Fig.~\ref{fig:alpha02_sigma_xx} but for absorptive, optical conductivity for right hand polarized light.}\label{fig:alpha02_sigma_plus}
\end{minipage}\qquad
\begin{minipage}[t]{.45\textwidth}
\includegraphics[width=\columnwidth]{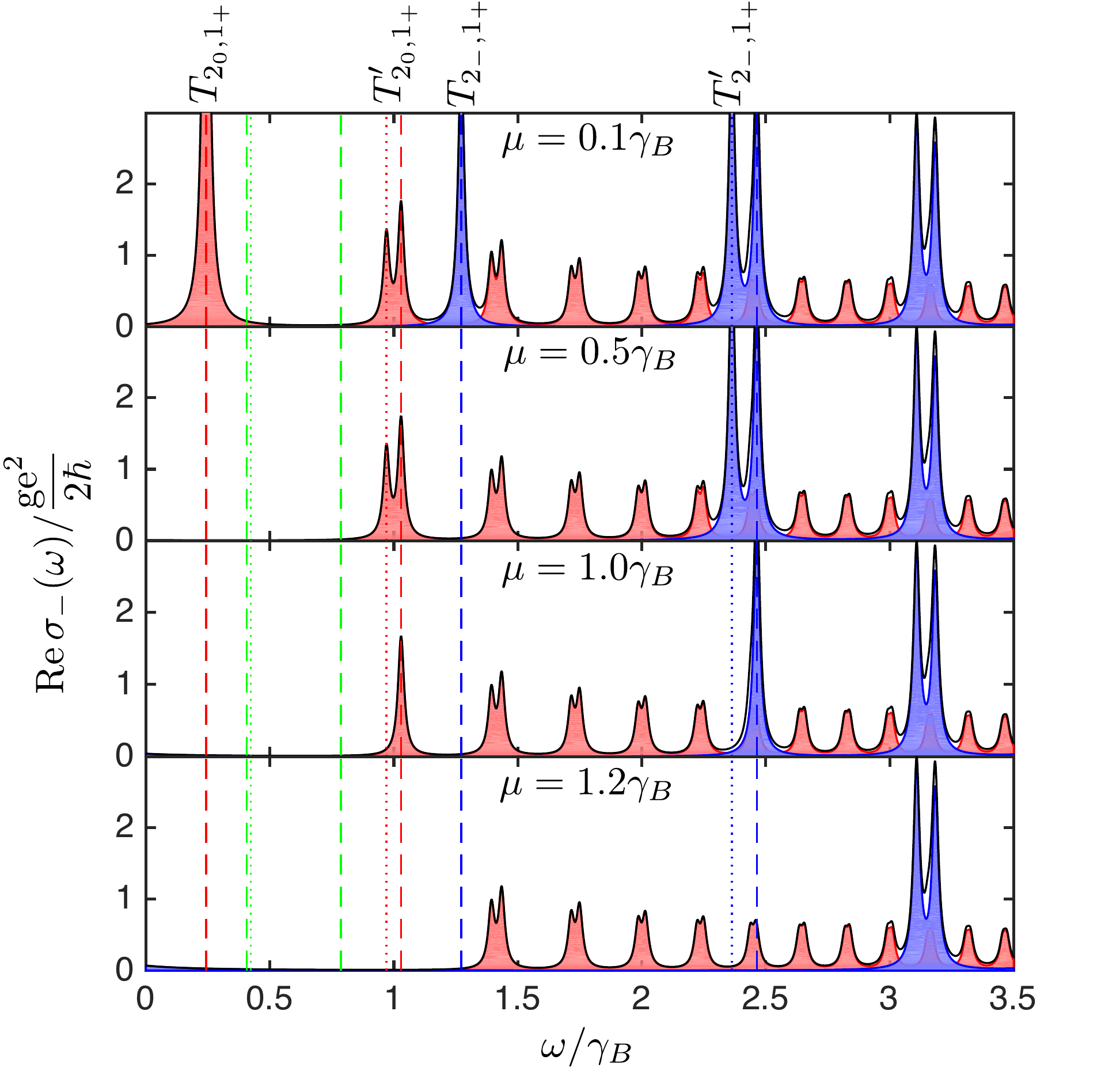}
\caption{(Color online) Same as Fig.~\ref{fig:alpha02_sigma_xx} but for the absorptive, optical conductivity for left hand polarized light.  The labels on the top have been changed to emphasize left directed transitions.}\label{fig:alpha02_sigma_minus}
\end{minipage}
\end{figure*}

We now turn to examining the magneto-optical response of the $\alpha$-$T_3$ lattice under a varying magnetic field, by making the magnitude of the magnetic field explicit in our calculations.  This allows us to connect more closely to experimental work where similar maps are an excellent tool for visualizing the LL structure and observing the magnetic field dependence of the observed transitions~\cite{jiang:2007,deacon:2007,shadowski:2006,plochocka:2008}.  

In Fig.~\ref{fig:mag_field_alphas}, we present a false-color map of the optical conductivity as a function of the square root of the magnetic field for four values of the parameter $\alpha$, including the two limiting cases of graphene and the dice lattice.  As one might expect from Eq.~\eqref{eq:energy1}, all of the observed transition energies depend on the applied magnetic field as $\sqrt{B}$ for all values of $\alpha$, with slopes that depend on the value of the parameter $\alpha$ and the LL index $n$.  We see the dominant cone-to-cone transition $T_{1_{\pm},2_+}$ for $\alpha=0$ and the dominant flat-band-to-cone transition $T_{0_0,1_+}$ for $\alpha=1$.  These are followed by additional transitions that decrease in intensity and become more tightly spaced with increasing $n$.

For $\alpha=0.5$ and $0.75$ we see additional structure in the spacing of transitions - in the form of doublets reminiscent of those in Fig.~\ref{fig:sigmaxx_alphas}.  In particular, the peak associated with the $T_{0_0,1_+}$ transition is split into its $K$ and $K^{\prime}$ valley counterparts and as a result, appears as two transitions of comparable intensity in this regime.  The overall doublet structure of the transitions is most apparent in the $\alpha=0.75$ colormap, where we can clearly see the pattern continue even for higher energy transitions.  
  
Next, we examine the magneto-optical response of the $\alpha$-$T_3$ lattice as a function of the chemical potential $\mu$.  We chose a value of the coupling parameter $\alpha$ such that both cone-to-cone and flat-band-to-cone transitions are well represented in our calculations.  As can be seen in Fig.~\ref{fig:sigmaxx_alphas}, $\alpha=0.25$ is an excellent representative case. 

In Fig~\ref{fig:snowshoe}, we present a snowshoe diagram~\cite{malcolm:2014} for $\alpha=0.25$.  Arrows represent transitions between LL, which are depicted as open circles connected by dashed and dotted lines in the $K$ and $K^{\prime}$ valleys, respectively.  The transitions shown are for a chemical potential of $\mu=0.1\gamma_B$.  Other chemical potentials of interest are also depicted in Fig.~\ref{fig:snowshoe}(a) as horizontal dashed lines, specifically $ 0.5\gamma_B, \,1.0\gamma_B,\, 1.2\gamma_B$.  These values were chosen such that for the lowest value of $\mu$, all positive LLs are above the chemical potential and for each successive value, $\mu$ is shifted past exactly one LL, either in the $K$ or $K^{\prime}$ valley.       

In Figs.~\ref{fig:alpha02_sigma_xx} through  \ref{fig:alpha02_sigma_minus} we present the magneto-optical conductivity curves for $\alpha=0.25$ including the absorptive part of the diagonal and the off-diagonal conductivities, $\rm{Re}\,\sigma_{xx}(\omega)$ and $\rm{Im}\,\sigma_{xy}(\omega)$; as well as the absorptive part of the conductivities for left and right hand polarized light, $\rm{Re}\,\sigma_{+}(\omega)$ and $\rm{Re}\,\sigma_{-}(\omega)$, respectively.  In these figures, vertical lines depict the photon energies of a number of transitions of interest.  The red, blue and green vertical lines represent the energies of flat-band-to-cone transitions, cone-to-cone interband transitions and cone-to-cone intraband transitions, respectively.  Transitions in the $K$ and $K^{\prime}$ valleys are shown with dashed and dot-dashed lines, respectively.  The subset of peaks that are affected by the first two shifts in chemical potential are labelled by representative transitions above the plots.  We shade the flat-band-to-cone response red, the cone-to-cone response blue, and denote the total optical response with a thin black curve. 

In Fig.~\ref{fig:alpha02_sigma_xx}, we examine the effect of shifting the chemical potential past the three lowest LLs of the $\alpha=0.25$ conductivity curve of Fig.~\ref{fig:sigmaxx_alphas}. Upon increasing the chemical potential above the first LL in the $K$ valley, a single red peak disappears, a blue peak is halved, and a new blue peak appears at an energy between the two original peaks.  Similarly, raising the chemical potential above the first LL in the $K^{\prime}$ valley results in the disappearance of a red peak, the halving of a blue peak, and the appearance of a new blue peak - this time at a lower energy than either of the original transitions.  

In both cases, the red peak that disappears is the lowest energy flat-band-to-cone transition ($T_{0_0,1_+}$ and $T_{0_0,1_+}^{\prime}$) for the respective valleys, the blue peak that is halved is the lowest energy cone-to-cone transition ($T_{1_-,2_+}$ and $T_{1_-,2_+}^{\prime}$) for the respective valleys, and the transition that appears is the intraband transition that crosses the new value of the chemical potential ($T_{1_+,2_+}$ and $T_{1_+,2_+}^{\prime}$) for the respective valleys.  Despite this, the action that takes place is not limited to the two lowest energy peaks.  In fact, higher energy peaks are affected by increases in chemical potential since the lowest energy peak from a particular transition type (ie, cone-to-cone or flat-band-to-cone) is not in general the lowest energy transition in the entire spectrum.  

We continue the trend in the bottom panel of Fig.~\ref{fig:alpha02_sigma_xx}, where the chemical potential is raised above the second LL in the $K$ valley.  We observe the disappearance of a red peak, and the disappearance of the second half of a blue peak.  The intraband transition that previously appeared is also replaced by one that crosses the new value of the chemical potential (ie, the transition $T_{1_+,2_+}$ is replaced by $T_{2_+,3_+}$).  As in the previous two shifts in $\mu$, peaks other the lowest energy ones are affected.

In contrast, for both graphene and the dice lattice, shifting the chemical potential past a single LL results in the halving or disappearance of the lowest energy cone-to-cone or flat-band-to-cone interband transition, respectively.  This may be accompanied by the disappearance of an intraband transition and the appearance of a new intraband transition at a lower energy.  Thus, for the limiting cases of $\alpha=0,1$ only the one or two lowest energy transitions are effected by a shift in the chemical potential.  In the hybrid system, multiple peaks are effected simultaneously, and these peaks are not in general the lowest energy peaks.  Thus, the effects of an increased chemical potential on higher energy transitions can serve as a signature of the hybrid system.  

We also note the difference in how peaks due to flat-band-to-cone versus cone-to-cone transitions disappear.  A flat-band-to-cone peak disappears completely with a single increase in chemical potential, since transitions that contribute to those peaks terminate at the same LL.  For cone-to-cone peaks, transitions that share the same energy terminate at LLs one index apart.  This results in a halving of a peak, followed by the disappearance of the second half of the peak upon blocking the next LL via another increase in chemical potential.  This difference between the response of cone-to-cone versus flat-band-to-cone peaks to increases in chemical potential introduces additional richness into the intermediate regime.  In this regime, some peaks vanish with a single shift of $\mu$, while others are only halved.   

In Fig.~\ref{fig:alpha02_sigma_xy} we plot the off-diagonal part of the absorptive optical conductivity.  Here, right directed transitions, denoted $T_{n,m}$ with $m=n+1$ are negative, and left directed transitions, denoted $T_{n,m}$ with $m=n-1$ are positive, as can be inferred from Eq.~\eqref{eq:matrix_elements}.  The snowshoe diagrams in Fig.~\ref{fig:snowshoe} depicts left and right directed transitions as arrows that point to the left and right, respectively.  Looking at the red and blue shaded peaks, we observe that the cone-to-cone transitions behave like those of graphene and the flat-band-to-cone transitions follow those of the dice lattice~\cite{malcolm:2014}.  In particular, peaks associated with flat-band-to-cone transitions are primarily positive, with the exception of the first mixed type transition.  Peaks associated with the cone-to-cone transitions are all negative, including the interband ones, due to the fact that the overlap functions cancel for the right and left directed transitions, and we observe only peaks that represent transitions from unpaired arrows.  

The off-diagonal conductivity nevertheless exhibits some unique features that are not present in the two limiting cases.  For chemical potentials above the lowest LL, there are twice as many negative peaks from cone-to-cone transitions as there were in the $S=1/2$ case, resulting from the difference in energies of the LLs in the $K$ and $K^{\prime}$ valleys.  This is notable as the number of peaks for $S=1/2$ is exactly two (See Fig. 5 in Ref~\cite{malcolm:2014}), and for the $\alpha$-$T_3$ model is exactly four.  Finally, we note the presence of a series of both positive and negative valued peaks which persist for larger values of chemical potential.  In contrast, graphene has only negative valued peaks, while the dice lattice exhibits a single negative valued peak followed by a series of positive ones. 
 
In Fig.~\ref{fig:alpha02_sigma_plus} and ~\ref{fig:alpha02_sigma_minus} we plot the absorptive part of the optical conductivity for right and left hand polarized light, respectively.  For right hand polarized light, we find only right directed transitions, $T_{n,m}$ with $m=n+1$, that are associated with arrows pointing to the right in the snowshoe diagram of Fig.~\ref{fig:snowshoe}.  Similarly, for left hand polarized light we find transitions represented by left facing arrows in Fig.~\ref{fig:snowshoe}, denoted $T_{n,m}$ with $m=n-1$.  The labelling of peaks in Figs.~\ref{fig:alpha02_sigma_xx} through ~\ref{fig:alpha02_sigma_plus} emphasizes right directed transitions, for convenience.  We reverse this labelling convention for Fig.~\ref{fig:alpha02_sigma_minus}, emphasizing instead the left directed transitions that are actually visible in that figure.  

For right and left polarized light, each peak in the conductivity curve is a result of a single transition.  Consequently, cone-to-cone transitions are no longer halved before disappearing, and instead completely disappear as the chemical potential is shifted past the relevant LL.  Also note that no intraband transitions exist for $\rm{Re}\,\sigma_{-}(\omega)$ since all such transitions have the form $T_{n_+,m_+}$ with $m=n+1$ and are only active for right hand polarized light.  

As before, in contrast to graphene and the dice lattice, we see transitions appear as doublets in the conductivity curves for polarized light.  Additionally, we see peaks that are not necessarily the lowest energy interband peaks effected by a single shift in chemical potential.  These are signatures of the hybrid system that persist with left and right hand polarized light.

\section{Hofstadter Butterfly}\label{sec:butterfly}

\begin{figure}[htb]
\centering
%\vspace{0.15cm}
 \includegraphics[width=0.8\columnwidth]{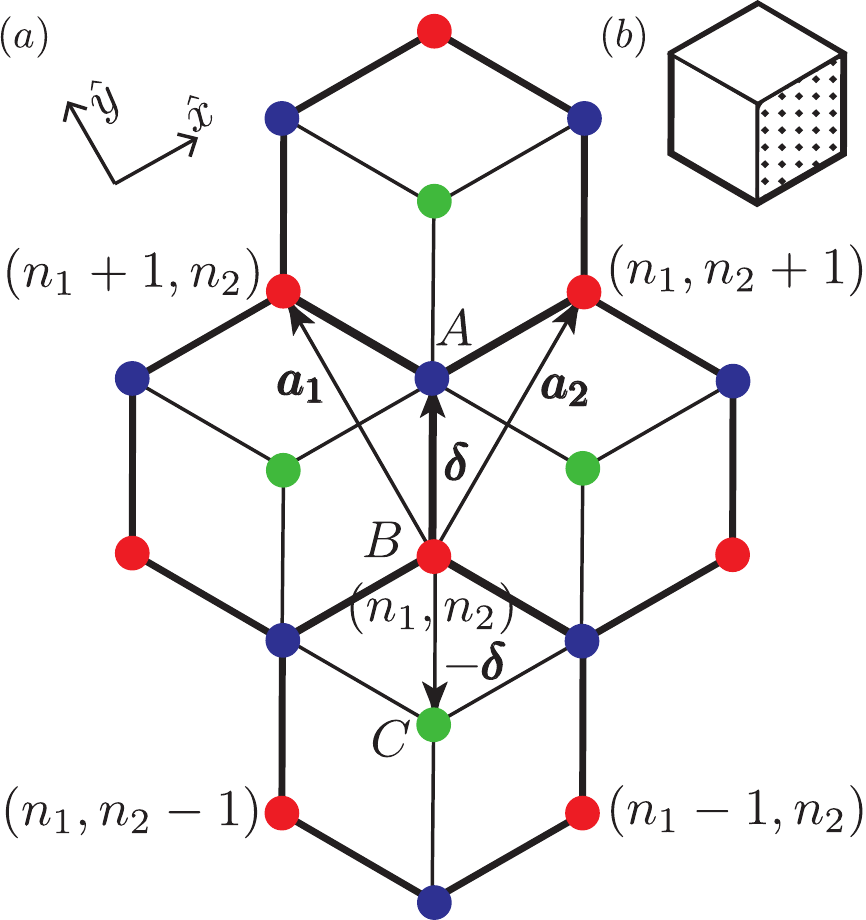}
\caption{(Color online) (a) The $\alpha$-$T_3$ lattice with three atoms per unit cell at site $A$, $B$ and $C$, represented by blue, red and green circles, respectively.  Primitive lattice vectors $\mathbf{a}_1$ and $\mathbf{a}_2$ and basis vector $\pmb{\delta}$ are depicted by arrows originating from lattice site $(n_1,n_2)$. (b) Schematic comparing the smallest plaquette for the $\alpha$-$T_3$ lattice for $\alpha\neq 1$ (dotted rhombus) versus $\alpha= 1$ (the entire hexagon).}
\label{fig:lattice}
\end{figure}

\begin{figure*}[ht!]
\centering
%\vspace{0.15cm}
 \includegraphics[width=0.9\textwidth]{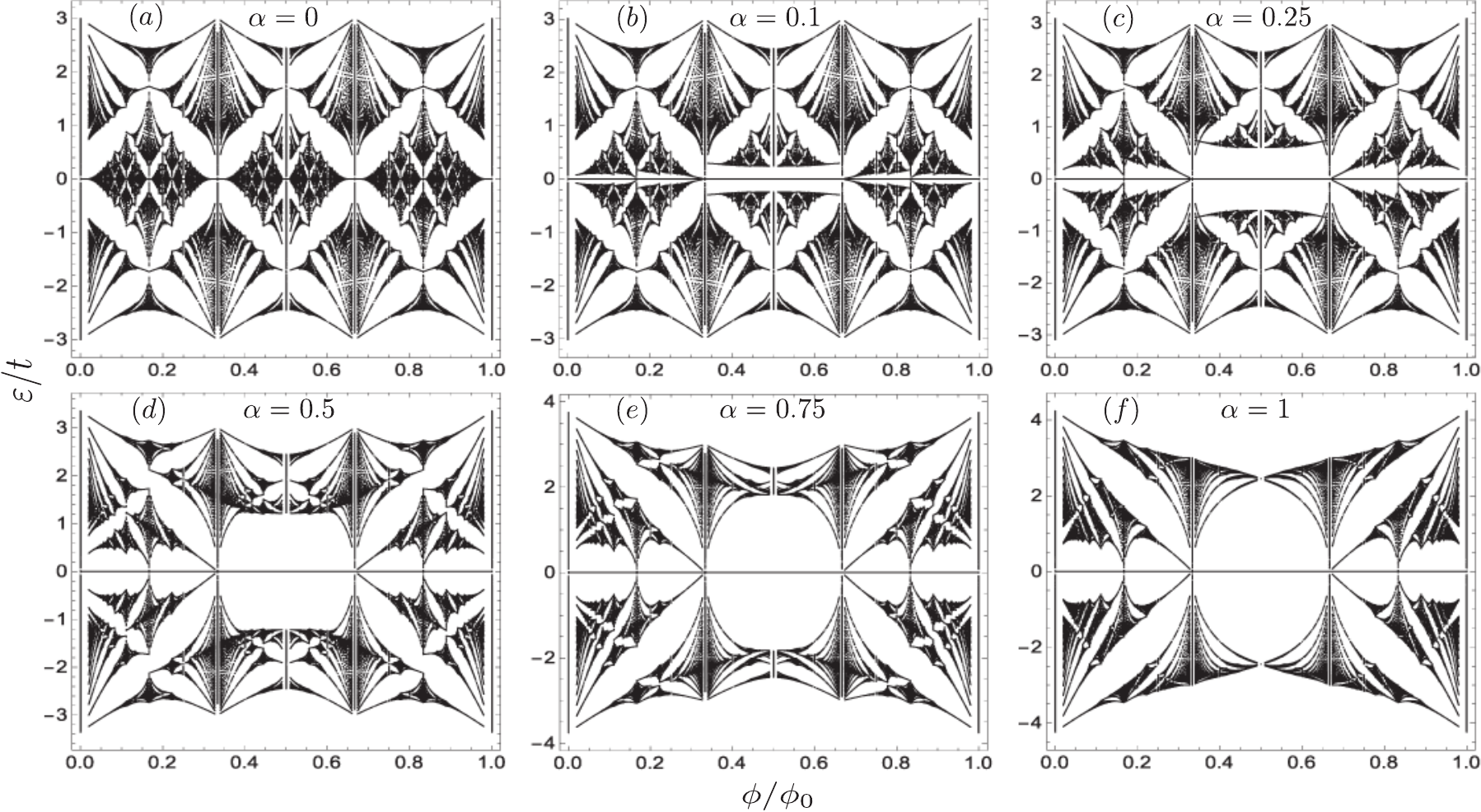}
\caption{Hofstadter butterflies for six representative $\alpha$ values calculated for $q$ up to $50$.}
\label{fig:butterflies}
\end{figure*}

In this section, we consider the $\alpha$-$T_3$ lattice in a perpendicular magnetic field and calculate the associated Hofstadter butterfly spectrum for the lattice.  

We begin by choosing primitive lattice vectors $\mathbf{a}_1=a\left(0,\sqrt{3}\right)$ and $\mathbf{a}_2=a\left(\frac{3}{2},\frac{\sqrt{3}}{2}\right)$ to span the lattice.  Here, $a$ is the interatomic distance, and we have chosen the vector $\mathbf{a}_1$ such that it lies in the $\hat{y}$ direction, for convenience.  We choose the $B$ sites as our lattice points, and use the basis vector $\pm\pmb{\delta}=\pm a\left(\frac{1}{2},\frac{\sqrt{3}}{2}\right)$ to access the atoms at sites $A$ and $C$, respectively (see Fig.~\ref{fig:lattice}). 

To denote the location of $A$, $B$ and $C$ atoms, we can now use the vector $\mathbf{R}_{n}$ which can be written in terms of the primitive lattice vectors and the basis vector as
\begin{align}\label{eq:r_vector}
	\mathbf{R}_{n_1,n_2,n_3} &= n_1\mathbf{a}_1+n_2\mathbf{a}_2+n_3\pmb{\delta}.
\end{align}
 
In the Landau gauge, the magnetic field in the $\hat{z}$ direction can be written $\mathbf{A}=Bx\hat{y}$.  Using the usual Peierls substitution, $\hbar k\rightarrow \hbar k -e\mathbf{A}/c$, the hopping $t$ picks up a phase $\theta_{n,m}$ in the presence of the field		
\begin{align}
	t_{n,m}\rightarrow t_{n,m} e^{-i\theta_{n,m}}.
\end{align}	
This phase can be calculated using 
\begin{align}
	\theta_{n,m}&=\frac{e}{\hbar c}\int_{\mathbf{R}_n}^{\mathbf{R}_m} \mathbf{A} \cdot \mathbf{dl}.
\end{align}	  
Between two arbitrary nearest neighbours located at $\mathbf{R}_n$ and $\mathbf{R}_m$ the phase is given by
\begin{align}
	\theta_{n,m}=& \frac{\pi B}{\phi_0}(\mathbf{R}_m-\mathbf{R}_n)_y(\mathbf{R}_n+\mathbf{R}_m)_x,
\end{align}
where $\phi_0=hc/e$ is the quantum flux, and the subscripts $x$ and $y$ refer to the $x$ and $y$ components of the respective vectors.
 
In the $\alpha$-$T_3$ lattice, atoms at sites $A$ and $C$ have three nearest neighbour atoms, while those at the $B$ sites have six nearest neighbours.  We can write down three coupled difference equations for the wave-functions at sites $A$, $B$ and $C$ with indices $(n_1,n_2)$
\begin{widetext}
\begin{align}\label{eq:three_diff}
\varepsilon\psi^B(n_1,n_2)&= t\Big[e^{-i\theta_{+}(n_2)}\psi^A(n_1,n_2)+e^{i\theta_{+}(n_2)}\psi^A(n_1-1,n_2)+ \psi^A(n_1,n_2-1)\Big]\nonumber\\	
&+ \alpha t\Big[e^{i\theta_{-}(n_2)}\psi^C(n_1,n_2)+e^{-i\theta_{-}(n_2)}\psi^C(n_1+1,n_2) 
+ \psi^C(n_1,n_2+1)\Big]\nonumber\\	
\varepsilon\psi^A(n_1,n_2)&= t\Big[e^{i\theta_{+}(n_2)}\psi^B(n_1,n_2) + e^{-i\theta_{+}(n_2)}\psi^B(n_1+1,n_2)
+ \psi^B(n_1,n_2+1)\Big]\nonumber\\	
\varepsilon\psi^C(n_1,n_2)&= \alpha t\Big[e^{-i\theta_{-}(n_2)}\psi^B(n_1,n_2) + e^{i\theta_{-}(n_2)}\psi^B(n_1-1,n_2) 
+ \psi^B(n_1,n_2-1)\Big]
\end{align}	
\end{widetext}
where $\theta_{\pm}(n_2)$ is the phase,  $\varepsilon$ is the energy and $\phi$ the elementary flux through a plaquette of the $\alpha$-$T_3$ lattice.  We have written the acquired phase $\theta_{n,m}(n_2)$ as $\theta_{\pm}(n_2)=\pi\frac{\phi}{\phi_0}(n_2\pm\frac{1}{6})$ in Eq.~\eqref{eq:three_diff} for the particular set of basis vectors we have chosen for the $\alpha$-$T_3$ lattice.  Here, the elementary flux $\phi=\frac{Ba^2 \sqrt{3}}{2}$, where $\frac{a^2 \sqrt{3}}{2}$ is the area of the smallest plaquette of the $\alpha$-$T_3$ lattice for $\alpha\neq0$ as depicted in Fig.~\ref{fig:lattice} (b).  Note that we have suppressed the third index $n_3$ in the wavefunctions of Eq.~\eqref{eq:three_diff}, since $n_3$ is always $1,0,-1$ for atoms at site $A$, $B$ and $C$, respectively (see Eq.~\eqref{eq:r_vector}).  

Upon combining the three difference equations from Eq.~\eqref{eq:three_diff} via substitution into the top equation one can obtain a single difference equation for $\psi^B(n_1,n_2)$ that is valid for $\varepsilon\neq 0$.  Taking into account the translational symmetry in the $\hat{y}$ direction due to the gauge choice~\cite{vidal:2001}, we can assume plane wave behaviour in this direction and look for solutions of the form
\begin{align}
	\psi^B(n_1,n_2)&=\varphi_{n_2}e^{ik_1n_1}\,\,\,\,\,\,\,\,k_1=\mathbf{a_1}\cdot\mathbf{k}=ak_y\sqrt{3}
\end{align}	
Simplification and some algebra yields a second order difference equation for $\varphi_{n_2}$
\begin{widetext}
\begin{align}\label{eq:butterfly}
\left[\varepsilon^2-3t^2(1+\alpha^2)\right]\varphi_{n_2}&=  2t^2\varphi_{n_2}\Big[\cos([6\pi\frac{\phi}{\phi_0}(n_2+\frac{1}{6})]-k_1)+\alpha^2\cos([6\pi\frac{\phi}{\phi_0}(n_2-\frac{1}{6})]-k_1)\Big]\nonumber\\
&+ 2t^2\varphi_{n_2-1}\Big[\cos([3\pi\frac{\phi}{\phi_0}(n_2-\frac{5}{6})]-\frac{k_1}{2})+\alpha^2\cos([3\pi\frac{\phi}{\phi_0}(n_2-\frac{1}{6})]-\frac{k_1}{2})\Big]\nonumber\\
&+ 2t^2\varphi_{n_2+1}\Big[\cos([3\pi\frac{\phi}{\phi_0}(n_2+\frac{1}{6})]-\frac{k_1}{2})+\alpha^2\cos([3\pi\frac{\phi}{\phi_0}(n_2+\frac{5}{6})]-\frac{k_1}{2})\Big]
\end{align}
\end{widetext}
It is easy to verify that Eq.~\eqref{eq:butterfly} reduces to the equation for the Hofstadter butterfly for the HCL and the dice lattice in the appropriate limits~\cite{rammal:1985,vidal:1998} of $\alpha=0$ and $\alpha=1$, respectively.  Rational values of $\frac{\phi}{\phi_0}=\frac{p}{q}$ make Eq.~\eqref{eq:butterfly} periodic.   Applying Bloch's theorem to take advantage of the periodicity yields a $q \times q$ eigenvalue equation for energy $\varepsilon$.  We solve this system of $q$ equations to obtain Hofstadter butterflies for the $\alpha$-$T_3$ lattice. 

In Fig.~\ref{fig:butterflies}, we show the Hofstadter spectra for six representative values of $\alpha$.  The spectra were calculated with a $q$ up to $50$ and plotted as a function of $\frac{\phi}{\phi_0}$.  The $\varepsilon=0$ solution that results from the non-dispersive flat band for all values of magnetic field is also included in the spectra though it is not given by Eq.~\eqref{eq:butterfly}.  As one might expect from Eq.~\eqref{eq:butterfly} and the symmetries of the $\alpha$-$T_3$ lattice, the Hofstadter butterfly spectra are symmetric about $\varepsilon=0$ and $\frac{\phi}{\phi_0}=\frac{1}{2}$, for all values of $\alpha$, with additional symmetries present in the $\alpha \rightarrow 0$ limit.

For the limiting case of $\alpha=1$, we obtain the Hofstadter butterfly spectrum of the dice lattice (see Fig.~\ref{fig:butterflies} (f)).  The spectrum has a highly degenerate eigenvalue resulting from the presence of the flat-band at $\varepsilon=0$ that carries $1/3$ of the total weight.  The spectrum contains a number of gaps, for example a large circular one near the center, which at $\frac{\phi}{\phi_0}=\frac{1}{2}$ is accompanied by a collapse of all the states to just three degenerate eigenvalues of $\varepsilon=0$ and $\varepsilon=\pm\sqrt{6}t$.  There are also a number of gapless bands, for example a large one at $\frac{\phi}{\phi_0}=\frac{1}{3}$ that stretches between $\varepsilon=\pm3t$.  A more detailed discussion of the Hofstadter butterfly spectrum for the dice lattice can be found in Refs.~\cite{vidal:1998,vidal:2001}.

For the other limiting case of $\alpha=0$, we obtain three repeats of the Hofstadter butterfly of graphene (see Fig.~\ref{fig:butterflies} (a)).  Focusing on a single repeat in the central region with $\frac{1}{3}<\Big|\frac{\phi}{\phi_0}\Big|<\frac{2}{3}$, the HCL Hofstadter butterfly is characterized by a set of gaps whose shape resembles the letter X, located at both positive and negative energies.  Repeats of this shape can be found throughout the complex fractal pattern of the HCL Hofstadter butterfly.  At $\frac{\phi}{\phi_0}=\frac{1}{2}$ there is a gapless band that stretches between $\varepsilon = \pm\sqrt{6}t$, in contrast to the three highly degenerate eigenvalues found for the dice lattice at the same flux.  

\begin{figure}[t!]
\centering
%\vspace{0.15cm}
 \includegraphics[width=\columnwidth]{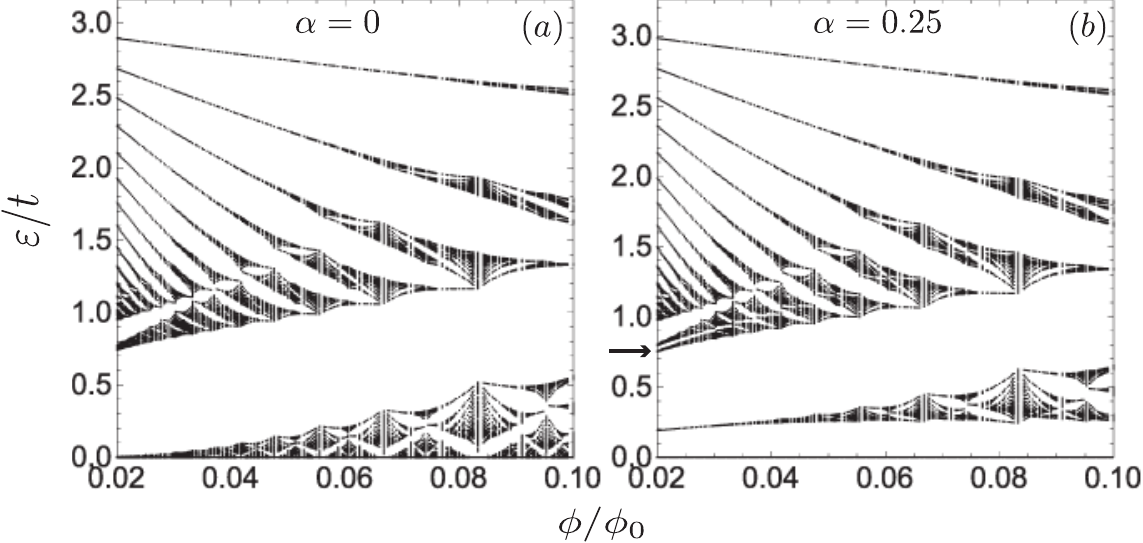}
\caption{A close-up of Hofstadter butterfly spectra for (a) $\alpha=0$ and (b) $\alpha=0.25$ for small fields calculated for $q$ up to $100$.  The arrow in (b) highlights the splitting of the LLs that is visible in the $\alpha=0.25$ butterfly.}
\label{fig:butterflies_zoom}
\end{figure}

Notably, we obtain three copies of the graphene Hofstadter butterfly spectrum, but only a single copy of the dice lattice one for the same range of $\frac{\phi}{\phi_0}$ in Fig.~\ref{fig:butterflies}.  This can be understood by looking at the diagram in Figure~\ref{fig:lattice}~(b) which contrasts the smallest plaquette that can be encircled by a semiclassical orbit for the HCL versus the $\alpha$-$T_3$ lattice.  Semiclassically, the smallest orbit an electron can make in the $\alpha$-$T_3$ lattice, with $\alpha\neq 0$, is along the edges of a rhombus with vertices $A$, $B$, $A$, $C$.  An example of such a rhombus is highlighted with dotted fill in Fig.~\ref{fig:lattice}~(b).  This rhombus has an area $\frac{\sqrt{3}a^2}{2}$.  In contrast, for $\alpha=0$, the atoms at the $C$ sites are inert, and cannot be part of a semiclassical orbit.  In this limit, the smallest orbit is the entire hexagon in Fig.~\ref{fig:lattice}~(b), which contains three copies of the rhombus, resulting in an area of $\frac{3\sqrt{3}a^2}{2}$.  Since $\phi=\frac{Ba^2 \sqrt{3}}{2}$, where $\frac{a^2 \sqrt{3}}{2}$ is the area of the smallest plaquette of the $\alpha$-$T_3$ lattice for $\alpha\neq0$, this results in three repeats of the Hofstadter butterfly for $\alpha=0$, where the area of the smallest plaquette is three times larger.

As $\alpha$ changes from $1$ to $0$ we observe the Hofstadter butterfly change its periodicity by a factor of three.  In the process, the large circular gap in the central region is squeezed from above and below, while the two side regions with $\Big|\frac{\phi}{\phi_0}\Big|>\frac{1}{3}$ symmetrically evolve to form two copies of the graphene Hofstadter butterfly spectrum.  During this process, a number of striking changes take place in the spectra.  For example, at $\frac{\phi}{\phi_0}=\frac{1}{2}$ the three highly degenerate eigenvalues of $\varepsilon=\pm\sqrt{3}t$ and $\varepsilon=0$ we observe for $\alpha=1$ become the large band that stretches between $\pm\sqrt{6}t$ for $\alpha=0$.  The most pronounced changes occur for smaller values of $\alpha$ as can be seen in Fig.~\ref{fig:butterflies} where much of the large central gap is still present for $\alpha=0.25$.

Recent seminal experiments in Moire superlattices have focused on observing the small field portion of the Hofstadter butterfly spectrum, since these regions are most readily accessible in the laboratory.  In Fig.~\ref{fig:butterflies_zoom} we highlight this portion of the Hofstadter butterfly spectrum for $\alpha=0$ and $\alpha=0.25$, for the $\alpha$-$T_3$ lattice.  For both $\alpha=0$ and $0.25$ we see a series of electron-like LLs that move up in energy with increasing $\phi$.  These are the LLs given by Eq.~\eqref{eq:energy1} that are formed in the cones located at the $K$ and $K^{\prime}$ points.  Additionally, we see hole-like LLs that move down in energy with increasing $\phi$.  These are accommodated in the hole pocket formed at the center of the hexagonal Brillouin zone~\cite{moon:2012}.  For $\alpha=0.25$, the splitting between the LLs in the $K$ and $K^{\prime}$ valley can be observed in the electron-like LLs.  This splitting is characterized by an unusually small spacing between LLs that interrupts the usual spacing observed between the remainder of the levels.  An example of this is shown by the arrow in Fig.~\ref{fig:butterflies_zoom} (b).
  
As the possibility of measuring the Hofstadter butterfly in graphene-like systems is starting to become a reality, it is appropriate to provide a characterization of this spectrum for the $\alpha$-$T_3$ model discussed here.

\section{Conclusions}\label{sec:conclusion}

In this paper we described the magneto-optical response and the Hofstadter butterfly spectrum of the $\alpha$-$T_3$ lattice.  We highlighted signatures of the intermediate regime between the pseudospin $S=1/2$ HCL and the pseudospin $S=1$ dice lattice.

In the magneto-optical conductivity, we noted a coexistence of the cone-to-cone transitions of graphene and flat-band-to-cone transitions of the dice lattice in the intermediate regime of the $\alpha$-$T_3$ model.  This was accompanied by a doubling of peaks associated with both transition types, a consequence of the inequivalent LL energies in the $K$ and $K^{\prime}$ valleys.  This interplay of the two transition types resulted in richness not observed in the two limiting cases, including anomalously sized peaks and doublets, as well as triplets of peaks.

Examining the magneto-optical response with a varying magnetic field $B$ showed $\sqrt{B}$ dependence for all transitions.  In the intermediate regime, a doublet structure in the peaks was again apparent, and in this case was manifest as pairs of transitions with comparable intensity.

For the HCL and the dice lattice, varying the chemical potential exclusively affects the lowest energy transitions of the magneto-optical conductivity curves.  In the intermediate regime of the $\alpha$-$T_3$ model, this action is not limited to the lowest energy peaks due to the richness of the mixing of cone-to-cone and flat-band-to-cone transitions.

Finally, we derived the difference equation required to calculate the Hofstadter butterfly spectrum for the intermediate regime of the $\alpha$-$T_3$ lattice.  This allowed us to describe the evolution of the Hofstadter spectrum as it changes its period by a factor of three can be observed in the intermediate regime.  Finally, we highlighted the low-field regime of the Hofstadter spectrum, as this is the regime most accessible for recent experiments in other lattices.  

\section{Acknowledgements}
We acknowledge J.P. Carbotte, G. Demand and J.D. Malcolm for useful discussions.  This work has been supported by the Natural Sciences and Engineering Research Council (NSERC) of Canada.

%\bibliography{AT3}

%merlin.mbs apsrev4-1.bst 2010-07-25 4.21a (PWD, AO, DPC) hacked
%Control: key (0)
%Control: author (0) dotless jnrlst
%Control: editor formatted (1) identically to author
%Control: production of article title (0) allowed
%Control: page (1) range
%Control: year (0) verbatim
%Control: production of eprint (0) enabled
%

\end{document}